%% file: main.tex
\documentclass[submission,copyright,creativecommons]{eptcs}
\providecommand{\event}{GandALF 2023} 

\usepackage{iftex}
\usepackage{lipsum}
\usepackage{float}
\usepackage{tikz}
\usetikzlibrary{arrows,automata,positioning,calc}
\usepackage{pgfplots}
\usepackage{verbatim}
\usepackage{MnSymbol}
\usepackage{amsmath}
\usepackage{amsfonts}
\usepackage{amsthm}
\usepackage{stmaryrd}
\usepackage{yfonts}
\usepackage{caption}
\usepackage{subcaption}
\usepackage{xcolor}
\usepackage{listings}
\usepackage{xspace}
\usepackage[linesnumbered,vlined,algoruled,commentsnumbered]{algorithm2e}
\SetKwInOut{Input}{Input}\SetKwInOut{Output}{Output}
\usepackage[capitalise,noabbrev,nameinlink]{cleveref}
\pgfplotsset{compat=1.17}

\lstdefinelanguage{LCGS}
{
  morekeywords={
    const, template, endtemplate, label, player
  },
  sensitive=false, 
  morecomment=[l]{//}, 
  morecomment=[s]{/*}{*/}, 
  morestring=[b]" 
}
\definecolor{eclipseBlue}{RGB}{42,0.0,255}
\definecolor{eclipseGreen}{RGB}{63,127,95}
\definecolor{eclipsePurple}{RGB}{147,0,85}
\ifpdf
  \usepackage[T1]{fontenc}        
\else
  \usepackage{breakurl}           
\fi

\newtheorem{theorem}{Theorem}[section]

\title{CGAAL: Distributed On-The-Fly ATL Model Checker with Heuristics\thanks{Funded by the VILLUM INVESTIGATOR project S4OS.}}
\author{
Falke B. Ø. Carlsen
\institute{Department of Computer Science,\\Aalborg University, Denmark}
\email{falkeboc@cs.aau.dk}
\and
Lars Bo P. Frydenskov
\institute{Department of Computer Science,\\Aalborg University, Denmark}
\email{larsbopark@gmail.com}
\and
Nicolaj Ø. Jensen
\institute{Department of Computer Science,\\Aalborg University, Denmark}
\email{noje@cs.aau.dk}
\and
Jener Rasmussen
\email{jener@jener.dk}
\and
Mathias M. Sørensen
\email{mathiasmehlsoerensen@gmail.com}
\and
Asger G. Weirsøe
\email{asger@weircon.dk}
\and
Mathias C. Jensen
\institute{Department of Computer Science,\\Aalborg University, Denmark}
\email{mcje@cs.aau.dk}
\and
Kim G. Larsen
\institute{Department of Computer Science,\\Aalborg University, Denmark}
\email{kgl@cs.aau.dk}
}
\def\titlerunning{CGAAL}
\def\authorrunning{F. Carlsen et al.}

\input{macros}

\begin{document}
\maketitle

\begin{abstract}
\input{Abstract}
{\color{white}\#fritfit}
\end{abstract}

\input{Sections/Introduction/Introduction}
\input{Sections/Definitions/Definitions}
\input{Sections/Model_Checking/Model_Checking}

\input{Sections/Tool_Overview/Tool_Overview}
\input{Sections/Evaluation/Evaluation}

\input{Sections/Conclusion}

\bibliographystyle{eptcs}
\bibliography{bib}


\end{document}

%% file: macros.tex
\NewDocumentCommand \thetool {}{\textsc{Cgaal}\xspace}

\NewDocumentCommand \prism {}{\textsc{Prism}\xspace}
\NewDocumentCommand \tapaal {}{\textsc{Tapaal}\xspace}

\NewDocumentCommand \czero {}{\textsc{CertainZero}\xspace}

\NewDocumentCommand \enforce m {\ensuremath{\llangle#1\rrangle}}
\NewDocumentCommand \despite m {\ensuremath{\llbracket#1\rrbracket}}
\NewDocumentCommand \eventually {} {\ensuremath{\Diamond}}
\NewDocumentCommand \atlnext {} {\ensuremath{\bigcirc}}
\NewDocumentCommand \invariantly {} {\ensuremath{\Box}}
\NewDocumentCommand \untilsymb {} {\ensuremath{\mathcal{U}}}
\NewDocumentCommand \until {m m} {\ensuremath{(#1 \; \untilsymb \; #2)}}

\NewDocumentCommand \R {} {\ensuremath{\mathbb{R}}}

\newcommand{\hyperedge}[4][180]{
     \draw (#2.#1)++(#1:0.7) edge (#2)
     edge[->] (#3)
     edge[->] (#4);
}

\newcommand{\hyperedgethree}[5][270]{
     \draw (#2.#1)++(#1:0.7) edge (#2)
     edge[->] (#3)
     edge[->] (#4)
     edge[->] (#5);
}

\NewDocumentCommand \testMSone {m m} {\ensuremath{MS_{#1}^{#2}\phi^{\bot\dagger}_1}}
\NewDocumentCommand \testMStwo {m m} {\ensuremath{MS_{#1}^{#2}\phi^{\bot}_2}}
\NewDocumentCommand \testMSthree {m m} {\ensuremath{MS_{#1}^{#2}\phi^{\top}_3}}

\NewDocumentCommand \testGPone {m} {\ensuremath{GG_{#1}\phi^{\top\dagger}_1}}
\NewDocumentCommand \testGPtwo {m} {\ensuremath{GG_{#1}\phi^{\top\dagger}_2}}
\NewDocumentCommand \testGPthree {m} {\ensuremath{GG_{#1}\phi^{\top\dagger}_3}}
\NewDocumentCommand \testGPfour {m} {\ensuremath{GG_{#1}\phi^{\top\dagger}_4}}
\NewDocumentCommand \testGPfive {m} {\ensuremath{GG_{#1}\phi^{\top\dagger}_5}}
\NewDocumentCommand \testGPsix {m} {\ensuremath{GG_{#1}\phi^{\bot\dagger}_6}}
\NewDocumentCommand \testGPseven {m} {\ensuremath{GG_{#1}\phi^{\top\dagger}_7}}

\NewDocumentCommand \testRCone {m} {\ensuremath{RC_{#1}\phi^{\bot}_1}}
\NewDocumentCommand \testRCtwo {m} {\ensuremath{RC_{#1}\phi^{\top}_2}}
\NewDocumentCommand \testRCthree {m} {\ensuremath{RC_{#1}\phi^{\top\dagger}_3}}
\NewDocumentCommand \testRCfour {m} {\ensuremath{RC_{#1}\phi^{\bot}_4}}

\NewDocumentCommand \testTTTone {} {\ensuremath{TTT\phi^{\top}_1}}
\NewDocumentCommand \testTTTtwo {} {\ensuremath{TTT\phi^{\top}_2}}

\NewDocumentCommand \testRPSone {} {\ensuremath{RPS\phi^{\bot\dagger}_1}}
\NewDocumentCommand \testRPStwo {} {\ensuremath{RPS\phi^{\bot}_2}}

\NewDocumentCommand \testMPone {} {\ensuremath{MP\phi^{\bot}_1}}
\NewDocumentCommand \testMPtwo {} {\ensuremath{MP\phi^{\top\dagger}_2}}
\NewDocumentCommand \testMPthree {} {\ensuremath{MP\phi^{\bot}_3}}

\NewDocumentCommand \testPAone {m} {\ensuremath{PA_{#1}\phi^{\top\dagger}_1}}
\NewDocumentCommand \testPAtwo {m} {\ensuremath{PA_{#1}\phi^{\top\dagger}_2}}
\NewDocumentCommand \testPAthree {m} {\ensuremath{PA_{#1}\phi^{\top}_3}}
\NewDocumentCommand \testPAfour {m} {\ensuremath{PA_{#1}\phi^{\bot}_4}}
\NewDocumentCommand \testPAfive {m} {\ensuremath{PA_{#1}\phi^{\top}_5}}

%% file: Abstract.tex
We present \thetool, our efficient on-the-fly model checker for alternating-time temporal logic (ATL) on concurrent game structures (CGS).
We present how our tool encodes ATL as extended dependency graphs with negation edges and employs the distributed on-the-fly algorithm by Dalsgaard et al.
Our tool offers multiple novel search strategies for the algorithm, including DHS which is inspired by PageRank and uses the in-degree of configurations as a heuristic, IHS which estimates instability of assignment values, and LPS which estimates the distance to a state satisfying the constituent property using linear programming.
CGS are input using our modelling language LCGS, where composition and synchronisation are easily described.
We prove the correctness of our encoding, and our experiments show that our tool \thetool is often one to three orders of magnitude faster than the popular tool \prism-games on case studies from \prism's documentation and among case studies we have developed.
In our evaluation, we also compare and evaluate our search strategies, and find that our custom search strategies are often significantly faster than the usual breadth-first and depth-first search strategies.

%% file: Sections/Introduction/Introduction.tex
\section{Introduction}\label{sec:introduction}
Software plays a large role in our everyday lives, making decisions, enabling efficient communication, ensuring safety, and many more critical tasks.
Furthermore, the complexity of the software and the decisions they have to make are ever-increasing due the to interconnectivity and reactive nature of modern systems.
These software systems mutually depend on, communicate with, and guide each other based on their collective and internal states.
Even a few interconnected systems that are not necessarily themselves too complex can give rise to an extremely complex system as a whole.
Safety-critical software that operates in contexts where errors could lead to human casualties or significant capital losses requires methods to verify their correctness.
Unfortunately, such methods are difficult to implement due to the sheer state-space explosion that arises from the inherent parallel composition of systems.

In this paper, we consider systems that can be expressed as discrete multiplayer games in which the actors can perform actions concurrently.
That is, in each configuration of the game, each player simultaneously chooses an action they wish to perform and the resulting decision vector then deterministically results in the next configuration of the game.
Such concurrent games are extremely expressive and can be used to model a variety of complex systems and lends themselves inherently, by their concurrent nature, to describing multi-actor systems.
The properties that we wish to verify on these systems are those that can be expressed by Alternating-time Temporal Logic (ATL)~\cite{atl}, a game theoretic extension of Computation Tree Logic (CTL)~\cite{Clarke1982ctl,Clarke2018Handbook}, that like CTL can be used to describe safety and liveness properties.
ATL extends CTL by replacing the existential and universal path quantifiers with so-called coalition quantifiers.
These coalition quantifiers describe properties of the game in which a coalition of players can either enforce some desired outcome or cannot avoid an outcome.
Of course, proof of satisfaction of such property is done by finding a strategy showing that the players can indeed enforce said property or by showing that no matter what strategy the players follow they cannot avoid said property.

One existing tool that can model check concurrent (stochastic) games is \prism-games~\cite{Kwiatkowska2020PrismGames}, an extension of the probabilistic model checker \prism~\cite{Parker2011Prism}.
\prism excels at stochastic models such as discrete- and continuous-time Markov chains, Markov decision processes, and (priced) probabilistic timed automata, for which it can verify various properties described in LTL , CSL, and probabilistic CTL*.
There are multiple verification engines in \prism, both symbolic and explicit-state.
The \prism-games extension focuses on stochastic games in which game-theoretic approaches are needed.
\prism-games can among other things synthesise strategies and reason about equilibria-based properties, where players may have distinct, but not necessarily conflicting objectives.
These properties are typically described with probabilistic ATL with rewards.
In version 3.0~\cite{Kwiatkowska2020PrismGames}, \prism-games added support for concurrent stochastic games.

In \cite{Liu1998SimpleLA}, Liu and Smolka present a global and a local algorithm to compute fixed-point boolean vertex assignments in dependency graphs with directed hyper-edges representing dependencies between vertex values.
The global algorithm has a better worst-case running time of the two, but the on-the-fly local algorithm only explores the graph as needed to compute the fixed-point assignment.
Many problems have since been encoded as dependency graphs.
Notably, model-checking problems for CTL can be encoded as dependency graphs and the algorithms by Liu and Smolka can be used to compute the satisfaction relation enabling model-checking using dependency graphs.
J.~F.~Jensen~et~al.~\cite{weightedCTL} show that weighted CTL can be encoded in dependency graphs as well and introduce symbolic edges to handle weights efficiently.
In~\cite{probabilisticCTL}, M.~C.~Jensen et~al. introduce symbolic dependency graphs, an encoding of probabilistic weighted CTL, and an implementation of a local and global algorithm.
A.~Dalsgaard et~al.~\cite{dalsgaard} present a distributed extension to the local algorithm to boost the performance and show how to model Petri Nets problems using CTL and dependency graphs. Furthermore, they extend dependency graphs by adding components and negation edges to solve formulae with negations.
An abstract dependency graph framework is presented by S.~Enevoldsen et~al. in \cite{senevoldsen2022abstractDG} where they generalise the various extensions of the dependency graphs with a general algorithm for any Noetherian partial order domain.
In the closely related work of~\cite{mcjensen2020patl}, S.~Enevoldsen~et~at.\ uses abstract dependency graphs to model check probabilistic ATL for weighted stochastic games.
In multiple works \cite{dalsgaard,probabilisticCTL} the local algorithm is found to be faster more often than the global algorithm in practice.

In this paper, we introduce the tool \thetool (Concurrent Games AALborg), a model checker implemented in Rust that allows for the verification of ATL properties on concurrent games.
\thetool works by encoding the model checking problem as an extended dependency graph and by implementing the on-the-fly local distributed algorithm by Dalsgaard et~al.~\cite{dalsgaard}.
The concurrent games are described and inputted to the tool using LCGS, our \prism-language inspired model language, and for specific ATL formulae, we can output a strategy describing how to ensure satisfaction given a positive result.
Furthermore, \thetool implements a variety of strategies for searching the problem state space: a heuristic inspired by PageRank~\cite{Page1998PageRank} and two heuristic search strategies that exploit the state vector representation of our model, alongside the usual breadth-first and depth-first approach.
We conduct experiments comparing our different search strategies with each other on several case studies and show that having more available compute threads often leads to speed-ups of 1-2 orders of magnitude.
We also perform experiments comparing our tool on a selection of case studies to the state-of-the-art \prism-games.
Though while \prism-games is a tool specialised for probabilistic games we find that our implementation often outperforms \prism-games and especially whenever we are not required to compute the entire fixed point in which case we are often 1-3 orders of magnitude faster.
Again, we emphasise that \prism-games is meant for probabilistic games and as such this is not a completely fair one-to-one comparison.

\paragraph{Outline.}
This paper is structured as follows.
\cref{sec:definitions} introduces the formal definitions of concurrent games and alternating-time temporal logic.
In \cref{sec:model_checking} we present how model checking of ATL properties in concurrent games can be encoded in extended dependency graphs.
We also give a short explanation of the \czero algorithm and our search strategies.
In \cref{sec:tool_overview} we give a short example of how to use \thetool and the LCGS language.
An evaluation of our tool and a comparison with \prism can be found in \cref{sec:evaluation}, and we conclude on our findings in \cref{sec:conclusion}.

%% file: Sections/Definitions/Definitions.tex
\section{Definitions}\label{sec:definitions}

We recall the definitions of concurrent games and alternating-time temporal logic.

\input{Sections/Definitions/Concurrent_Games}
\input{Sections/Definitions/Alternating-time_Temporal_Logic}

%% file: Sections/Definitions/Concurrent_Games.tex
\paragraph{Concurrent Games}\label{sec:concurrent_games}

A \textit{concurrent game structure} (CGS) is a tuple $S = \langle k, Q, \Pi, \pi, d, \delta \rangle$ where:
\begin{itemize}\itemsep0em
    \item $k\geq 1$ is a natural number of players. We identify the players with the numbers $1,\dotsc,k$.
    \item $Q$ is a finite set of states.
    \item $\Pi$ is a finite set of atomic propositions, also called labels.
    \item A set $\pi(q)\subseteq\Pi$ of propositions are true at $q\in Q$. The function $\pi$ is called the \textit{labelling function}.
    \item For each player $a\in\{1,\dotsc,k\}$ and each state $q\in Q$, a natural number $d_a (q)\geq 1$ of moves are available to player $a$ at state $q$. The moves of player $a$ at state $q$ are identified with the numbers $1,\dotsc,d_a (q)$. 
    
    A move vector at $q\in Q$ is a tuple $v=\langle j_1 ,\dotsc, j_k\rangle$ such that $1\leq j_a \leq d_a (q)$ for each player $a\in\{1,\dotsc,k\}$. Additionally, given a state $q\in Q$, we write $D(q)$ for the set $\{1,\dotsc,d_1 (q)\} \times\dotsb\times \{1,\dotsc,d_k (q)\}$ of all move vectors possible at $q$. The function $D$ is called the \textit{move function}.
    
    \item For each state $q\in Q$ and each move vector $v=\langle j_1 ,\dotsc, j_k\rangle\in D(q)$, we have that $\delta(q, v)\in Q$ is the resulting state when each player $a\in\{1,\dotsc,k\}$ chooses move $j_a$ in the state $q$. The function $\delta$ is called the \textit{transition function}.
\end{itemize}

A state $q'$ is a \textit{successor} to $q$ if and only if there exists a transition from $q$ to $q'$, i.e. there exists a move vector $v\in D(q)$ such that $\delta(q,v)=q'$. A \textit{computation} is an infinite sequence $\lambda=q_0,q_1,q_2,\dotsc$ of states, such that for all $i\geq 0$, the state $q_{i+1}$ is a successor of $q_i$. Given a computation $\lambda$ and a position $i\geq 0$, we use the following notations:
\begin{align*}
    \lambda[i] & \quad \text{The $i$th state of the computation of $\lambda$} \\
    \lambda[0,i] & \quad \text{The finite prefix $q_0,q_1,\dotsc,q_i$ of $\lambda$} \\
    \lambda[i,\infty] & \quad \text{The infinite suffix $q_i, q_{i+1},q_{i+2},\dotsc$ of $\lambda$}
\end{align*}

A computation starting in the state $q$ is called a $q$-computation.

\paragraph{Game strategies}\label{sec:game_strategies}

Consider a concurrent game structure $S = \langle k, Q, \Pi, \pi, d, \delta \rangle$ over the set $\Sigma=\{1,\dotsc,k\}$ of players. A \textit{strategy} for a player $a\in\Sigma$ is a function $z_a:Q\to \mathbb{N}$ that maps every state $q\in Q$ to a natural number, such that $z_a(q)\leq d_a(q)$. In other words, the strategy $z_a$ describes how player $a$ chooses their move in each state. A strategy $z_a$ for a player $a\in\Sigma$ induces a set of computations that the player $a$ can enforce. Given a state $q\in Q$, a set $A\subseteq\Sigma$ of players, and a set $Z_A=\{z_a\}_{a\in A}$ of strategies, one strategy for each player in $A$, we define the outcomes of the strategies $Z_A$ from $q$, denoted $out(q,Z_A)$, to be the set of $q$-computations that the players of $A$ enforce following the strategies of $Z_A$. That is, a computation $\lambda=q_0,q_1,q_2,\dotsc$ is in $out(q,Z_A)$ if $q_0=q$ and for all positions $i\geq 0$, there is a move vector $v=\langle j_1, \dotsc,j_k\rangle\in D$ such that
\begin{enumerate}
    \item $j_a=z_a(q_i)$ for all players $a\in A$, and
    \item $\delta(q_i, v)=q_{i+1}$.
\end{enumerate}

The set $\mathcal{Z}^A=\{\{z_a\}_{a\in A}\mid z_a \text{ is a strategy for the player } a\}$ contains all sets of strategies, that contain one strategy for each player in $A\subseteq\Sigma$.

%% file: Sections/Definitions/Alternating-time_Temporal_Logic.tex
\paragraph{Alternating-time Temporal Logic}\label{sec:alternating-time_temporal_logic}

The \textit{alternating-time temporal logic} (ATL) \cite{atl}  is defined with respect to a finite set $\Pi$ of \textit{propositions} and a finite set $\Sigma=\{1, ..., k\}$ of players. An ATL formula is given by the abstract syntax:
\[
    \phi ::= p \mid \neg\phi \mid \phi_1\lor\phi_2 \mid \llangle A\rrangle\bigcirc\phi \mid \llangle A\rrangle(\phi_1\mathcal{U}\phi_2) \mid \llbracket A\rrbracket(\phi_1\mathcal{U}\phi_2)
\]
\noindent where $p\in\Pi$ is a proposition and $A\subseteq\Sigma$ is a set of players. The operators $\llangle\cdot\rrangle$ and $\llbracket\cdot\rrbracket$ are \textit{path quantifiers}. We will refer to them as "enforce" and "despite", respectively. The $\bigcirc$ ("next") and $\mathcal{U}$ ("until") are \textit{temporal operators}. Additional temporal operators like $\Diamond$ ("eventually") and $\Box$ ("invariant") are derived as usual.


Given a state $q$ of a game structure $S$, we write $q\vDash\phi$ to indicate that the state $q$ satisfies the property described by $\phi$. The satisfactory relation $\vDash$ is defined inductively:

\begin{itemize}\itemsep0em
    \item $q\vDash p$ iff $p\in\pi(q)$.
    \item $q\vDash\neg\phi$ iff $q\not\vDash\phi$.
    \item $q\vDash\phi_1\lor\phi_2$ iff $q\vDash\phi_1$ or $q\vDash\phi_2$.
    \item $q\vDash\llangle A\rrangle\bigcirc\phi$ iff there exists a set $Z_A\in\mathcal{Z}^A$ of strategies, such that for all computations $\lambda\in out(q,Z_A)$, we have $\lambda[1]\vDash\phi$.
    \item $q\vDash\llangle A\rrangle(\phi_1\mathcal{U}\phi_2)$ iff there exists a set $Z_A\in\mathcal{Z}^A$ of strategies, such that for all computation $\lambda\in out(q,Z_A)$ there exists a position $i\geq 0$, such that $\lambda[i]\vDash\phi_2$ and for all positions $0\leq j < i$, we have $\lambda[j]\vDash\phi_1$.
    \item $q\vDash\llbracket A\rrbracket(\phi_1\mathcal{U}\phi_2)$ iff for all sets $Z_A\in\mathcal{Z}^A$ of strategies, we have that there exists a computation $\lambda\in out(q,Z_A)$, such that there exists a position $i\geq 0$, such that $\lambda[i]\vDash\phi_2$ and for all positions $0\leq j < i$, we have $\lambda[j]\vDash\phi_1$.
\end{itemize}

%% file: Sections/Model_Checking/Model_Checking.tex
\section{Model Checking}\label{sec:model_checking}

In order to check if a CGS satisfies an ATL property, \thetool encodes the problem as an extended dependency graph and finds a fixed-point assignment describing the satisfaction relation.

\input{Sections/Model_Checking/Extended_Dependency_Graphs}
\input{Sections/Model_Checking/Assignments}
\input{Sections/Model_Checking/Encoding_of_ATL_as_EDG}
\input{Sections/Model_Checking/Certain_Zero}
\input{Sections/Model_Checking/Search_Strategies/Search_Strategies}

%% file: Sections/Model_Checking/Extended_Dependency_Graphs.tex
\paragraph{Extended Dependency Graphs}\label{sec:extended_dependency_graphs}

An \textit{extended dependency graph} (EDG) is a tuple $G = \langle C, E, N\rangle$ where $C$ is a finite set of configurations (vertices), $E\subseteq C \times\mathcal{P}(C)$ is a finite set of hyper-edges, and $N \subseteq C \times C$ is a finite set of negation edges. 

For a hyper-edge $e = \langle c,T\rangle \in E$ we call $c$ the source configuration and $T \subseteq C$ the set of target configurations. Similarly, $c$ is called the source configuration of the negation edge $\langle c, c'\rangle\in N$. We write $c \rightarrow c'$ if there exists an edge $\langle c,T\rangle \in E$ such that $c' \in T$ and $c \dashrightarrow c'$ if $\langle c,c'\rangle\in N$. Furthermore, we write $c \rightsquigarrow c'$ if $c \rightarrow c'$ or $c \dashrightarrow c'$.
An EDG $G = \langle C,E,N\rangle$ is negation safe if there are no $c,c' \in C$ such that $c \dashrightarrow c'$ and $c' \rightsquigarrow^* c$. In what follows, we consider only negation-safe EDGs. Let $dist: C \rightarrow \mathbb{N}_0$ be the maximum number of negation edges throughout all paths starting in a configuration $c \in C$, inductively defined as:
\begin{align}
    dist(c) = \max\{dist(c'')+1 \mid c',c'' \in C\text{ and }c \rightarrow^* c' \dashrightarrow c'' \}
\end{align}
By convention $\max\emptyset = 0$. We define the $dist(G)$ of an EDG $G$ as $dist(G) = \max_{c \in C}(dist(c))$.
A component $K_i$ of an EDG $G$, where $i \in \mathbb{N}_0$, is a subgraph induced on $G$ by the sets
\begin{itemize}\itemsep0em
    \item $C_i = \{c \in C \mid dist(c) \leq i \}$
    \item $E_i = \{\langle c, T\rangle \in E \mid dist(c) \leq i \}$
    \item $N_i = \{\langle c, c'\rangle \in N \mid dist(c) \leq i \}$
\end{itemize}
denoting the set of configurations, hyper-edges and negation edges respectively in each respective component. By definition, the component $K_0$ has no negation edges.

%% file: Sections/Model_Checking/Assignments.tex
\paragraph{Assignments}\label{sec:assignments}

An assignment $\alpha: C \to \{0, 1\}$ is a function that assigns boolean values to configurations of an EDG $G=\langle C, E, N\rangle$, where 0 and 1 represent false and true, respectively.
We define $\alpha_0$ as the zero assignment where $\alpha_0(c)=0$ for all $c \in C$.
We assume a component-wise ordering $\sqsubseteq$ on assignments such that we have $\alpha \sqsubseteq \alpha'$ whenever $\alpha(c) \leq \alpha'(c)$ for all $c \in C$. The set of all assignments of an EDG $G$ is denoted by $\mathcal{A}^G$ and $\langle\mathcal{A}^G, \sqsubseteq\rangle$ is a complete lattice.

By Knaster and Tarski's theorem, we can find a minimum fixed point on the complete lattice for any monotonic function. The minimum fixed-point assignment $\alpha_{\min}$ of an EDG $G$, denoted as $\alpha^G_{\min} = \alpha^{K_{dist(G)}}_{\min}$ is defined inductively on the components $K_0, K_1, \dotsc, K_{dist(G)}$ of $G$. For all $0 \leq i \leq dist(G)$, we define $\alpha^{K_i}_{\min}$ to be the minimum fixed-point assignment of the monotonic function $F_i : \mathcal{A}^{C_i} \to \mathcal{A}^{C_i}$ where

\begin{align}\label{eq:assignment_mono_func}
    F_i(\alpha)(c) = \alpha(c) \lor \Bigg[\bigvee_{\substack{\langle c,T\rangle \in E_i}} \bigwedge_{\substack{c' \in T}} \alpha(c')\Bigg] \lor \Bigg[ \bigvee_{\substack{\langle c,c'\rangle \in N_i}} \neg \alpha^{K_{i-1}}_{\min} (c') \Bigg]
\end{align}

By convention, the conjunction of $\emptyset$ is true and the disjunction of $\emptyset$ is false. In the component $K_0$ there are no negation edges, which means the last clause in the disjunction is false for $K_0$.
We will refer to the repeated use of the $F_i$ as the \textit{global algorithm} because it requires building and iterating over the entire EDG.
By \cref{eq:assignment_mono_func} each configuration in the EDG is comparable to a boolean formula in disjunctive normal form where each clause is another configuration or a negation of another configuration.
Clearly, this makes EDGs an expressive structure and we shall now encode ATL model checking in an EDG.

%% file: Sections/Model_Checking/Encoding_of_ATL_as_EDG.tex
\subsection{Encoding of ATL in an EDG}\label{sec:encoding_of_atl_as_edg}

We first establish some definitions related to subsets of move vectors induced by a coalition of players.

\paragraph{Partial Moves.}
Given a CGS $S$ with the set of states $Q$ and $k$ players, we use $V_1\times\dotsb\times V_k=\mathcal{V}\subseteq D(q)$ where $q\in Q$ to denote \textit{a partial} move in $q$, where zero or more of the players' moves are fixed, i.e. their set of possible moves contain a single number. For instance, if $\mathcal{V}=\{2\}\times\{1, 2, 3\}$ in a 2-player game, then player 1 has chosen move 2 while player 2 is still free to choose from 1, 2, or 3. Furthermore, we use the notation $\mathcal{V}[a\mapsto j]$ for a variant $\mathcal{V}'$ of a partial move $\mathcal{V}=V_1\times\dotsb\times V_k$ where player $a$ chooses move $j$. More precisely
\begin{align}
    \mathcal{V}[a\mapsto j] = V'_1\times\dotsb\times V'_k\quad\text{where } 1\leq i\leq k \text{ and } V'_i = \begin{cases} \{j\} & \text{if } i = a \\ V_i & \text{otherwise} \end{cases}
\end{align}

With this notation in mind, we define the function $pmoves$. The function $pmoves$ gives the set of all possible partial moves that follow from the players of the set $A\subseteq\Sigma$ making a combination of moves in the state $q\in Q$:

\begin{align}
    pmoves(q, A) = \{ D(q)[a \mapsto j_a][b \mapsto j_b]\dotsb \mid \{a,b,\dotsc\} = A \text{ and } \forall i\in A . 1\leq j_i\leq d_i(q)\}
\end{align}

For instance, if $D(q)=\{1,2\}\times\{1,2\}$ in a 2-player game and $A=\{1\}$. Then $pmoves(q, A)=\{\{1\}\times\{1,2\}, \{2\}\times\{1,2\}\}$ contains two partial moves. One where player 1 chooses action 1, and another where player 1 chooses action 2. In other words, the $pmoves$ function constructs the partial moves that the set $A$ of players can choose from when \textit{working together} in the state $q$. If $A$ chooses $\mathcal{V}\in pmoves(q, A)$, then $\mathcal{V}$ contains all move vectors that result from the remaining players $\Sigma\backslash A$ also making a choice.

We define a \textit{partial transition function} $\Delta$ that produces a set of possible successor states given a state $q\in Q$ and a partial move $\mathcal{V}$. That is
\begin{align}
    \Delta(q, \mathcal{V}) = \{s \mid v\in\mathcal{V} \text{ and }\delta(q, v)=s\} \quad\text{where } \mathcal{V}\subseteq D(q)
\end{align}

With EDGs, assignments, and partial moves introduced, we can now define how the satisfaction relation $\vDash$ can be encoded as an EDG.

\begin{figure}
    \centering
    \begin{tikzpicture}
        \tikzstyle{conf}=[rectangle,draw]
        \node[conf] (root) at (0, 0) {$\langle q, \textbf{true}\rangle$};
        \node (p) at (0, -1.3) {$\emptyset$};
        \draw[->] (root) -- (p);
    \end{tikzpicture}\hspace{2cm}%
    \begin{tikzpicture}
        \tikzstyle{conf}=[rectangle,draw]
        \node[conf] (root) at (0, 0) {$\langle q, \neg\phi\rangle$};
        \node[conf] (phi) at (0, -1.3) {$\langle q, \phi\rangle$};
        \draw[dashed,->] (root) -- (phi);
    \end{tikzpicture}\hspace{2cm}%
    \begin{tikzpicture}
        \tikzstyle{conf}=[rectangle,draw]
        \node[conf] (root) at (0, 0) {$\langle q, p\rangle$};
        \node (p) at (0, -1.3) {$\emptyset$};
        \node[right=2mm of root] {if $p\in\pi(q)$};
        \draw[->] (root) -- (p);
    \end{tikzpicture}
    \begin{tikzpicture}
        \tikzstyle{conf}=[rectangle,draw]
        \node[conf] (root) at (0, 0) {$\langle q, p\rangle$};
        \node at (0, -0.7) {};
        \node[right=2mm of root] {if $p\notin\pi(q)$};
    \end{tikzpicture}
    \caption{EDG encoding of true, negation, and atomic proposition}
    \label{fig:dg_p_true_neg}
    
    \vspace{0.5cm}
    \begin{tikzpicture}
        \tikzstyle{conf}=[rectangle,draw]
        \node[conf] (root) at (0, 0) {$\langle q, \phi_1 \lor \phi_2\rangle$};
        
        \node[conf] (l) at (-1, -1) {$\langle q,\phi_1\rangle$};
        
        \node[conf] (r) at (1, -1) {$\langle q,\phi_2\rangle$};
        
        \draw[->] (root) -- (l);
        \draw[->] (root) -- (r);
    \end{tikzpicture}
    \caption{EDG encoding of disjunction}
    \label{fig:dg_or}
    
    \vspace{0.5cm}
    \begin{tikzpicture}[yscale=0.8]
        \tikzstyle{conf}=[rectangle,draw]
        \node[conf] (root) at (0, 0) {$\langle q, \llangle A\rrangle\bigcirc\phi\rangle$};
        
        \node[conf] (q1) at (-3, -2) {$\langle q_1, \phi\rangle$};
        \node at (-2, -2) {$\dotsb$};
        \node[conf] (qn) at (-1, -2) {$\langle q_n,\phi\rangle$};
        \node at (0, -2) {$\dotsb$};
        \node[conf] (s1) at (1, -2) {$\langle s_1,\phi\rangle$};
        \node at (2, -2) {$\dotsb$};
        \node[conf] (sm) at (3, -2) {$\langle s_m,\phi\rangle$};
        
        \hyperedge[220]{root}{q1}{qn}
        \hyperedge[320]{root}{s1}{sm}
        
        \node[right=0.3 of root] {where $\{\{q_1,\dotsc,q_n\},\dotsc,\{s_1,\dotsc,s_m\}\}=\{\Delta(q, \mathcal{V})\mid \mathcal{V}\in pmoves(q, A)\}$};
    \end{tikzpicture}
    \caption{EDG encoding of next}
    \label{fig:dg_next}

    \vspace{0.5cm}
    \begin{tikzpicture}[yscale=0.95]
        \tikzstyle{conf}=[rectangle,draw]
        \node[conf] (root) at (-1, 0) {$\langle q,\llangle A\rrangle(\phi_1\mathcal{U}\phi_2)\rangle$};
        
        \node[conf] (root1) at (-4, -1) {$\langle q,\phi_2\rangle$};
        \draw[->] (root) -- (root1);
        
        \node[conf] (root2) at (-3.4, -2) {$\langle q,\phi_1\rangle$};
        \node[conf] (q1) at (-2.5, -3) {$\langle q_1,\llangle A \rrangle(\phi_1\mathcal{U}\phi_2)\rangle$};
        \node at (-0.6, -3) {$\dotsb$};
        \node[conf] (qn) at (1.3, -3) {$\langle q_n,\llangle A \rrangle(\phi_1\mathcal{U}\phi_2)\rangle$};
        \node at (3.2, -3) {$\dotsb$};
        \node[conf] (s1) at (5.1, -3) {$\langle s_1,\llangle A \rrangle(\phi_1\mathcal{U}\phi_2)\rangle$};
        \node at (7, -3) {$\dotsb$};
        \node[conf] (sm) at (8.9, -3) {$\langle s_m,\llangle A \rrangle(\phi_1\mathcal{U}\phi_2)\rangle$};
        
        \hyperedgethree[220]{root}{root2}{q1}{qn}
        \hyperedgethree[320]{root}{root2}{s1}{sm}
        
        \node[right=0.3 of root] {where $\{\{q_1,\dotsc,q_n\},\dotsc,\{s_1,\dotsc,s_m\}\}=\{\Delta(q, \mathcal{V})\mid \mathcal{V}\in pmoves(q, A)\}$};
    \end{tikzpicture}
    \caption{EDG encoding of enforce until}
    \label{fig:dg_until}
    
    \vspace{0.5cm}
    \begin{tikzpicture}[yscale=0.95]
        \tikzstyle{conf}=[rectangle,draw]
        \node[conf] (root) at (-1, 0) {$\langle q,\llbracket A\rrbracket(\phi_1\mathcal{U}\phi_2)\rangle$};
        
        \node[conf] (root1) at (-4, -1) {$\langle q,\phi_2\rangle$};
        \draw[->] (root) -- (root1);
        
        \node[conf] (root2) at (-3.4, -2) {$\langle q,\phi_1\rangle$};
        \node[conf] (v1) at (-0.5, -2) {$\langle q,\mathcal{V}_1,\llbracket A \rrbracket(\phi_1\mathcal{U}\phi_2)\rangle$};
        \node at (1.5, -2) {$\dotsb$};
        \node[conf] (vn) at (3.5, -2) {$\langle q,\mathcal{V}_n,\llbracket A \rrbracket(\phi_1\mathcal{U}\phi_2)\rangle$};
        
        \hyperedgethree[270]{root}{root2}{v1}{vn}
        
        \node[right=of root] {where $\{\mathcal{V}_1,\dotsc,\mathcal{V}_n\} = pmoves(q, A)$};
    \end{tikzpicture}
    \caption{EDG encoding of despite until}
    \label{fig:dg_anti_until}
    
    \vspace{0.5cm}
    \begin{tikzpicture}[yscale=0.8]
        \tikzstyle{conf}=[rectangle,draw]
        \node[conf] (root) at (0, 0) {$\langle q, \mathcal{V}, \phi \rangle$};
        
        \node[conf] (q1) at (-1, -1.5) {$\langle q_1,\phi\rangle$};
        \node at (0, -1.5) {$\dotsb$};
        \node[conf] (qn) at (1, -1.5) {$\langle q_n,\phi\rangle$};
        
        \draw[->] (root) -- (q1);
        \draw[->] (root) -- (qn);
        
        \node[right=of root] {where $\{v_1,\dotsc,v_n\} = \mathcal{V}$ and $q_i=\delta(q,v_i)$ for $1\leq i \leq n$};
    \end{tikzpicture}
    \caption{EDG encoding of partially moved despite until}
    \label{fig:dg_partial}
\end{figure}

\paragraph{Encoding.}
Given a CGS $S$, a state $q$ of $S$, and an ATL state formula $\phi$ we now construct an EDG where every configuration is either a pair of a state and a formula, or a triple of a state, partial move, and a formula. The triples represent partially evaluated states, where some players' moves are already set. Starting from the initial pair $\langle q, \phi\rangle$, the dependency graph is constructed according to \cref{fig:dg_p_true_neg,fig:dg_or,fig:dg_next,fig:dg_until,fig:dg_anti_until,fig:dg_partial}. The figures show which outgoing edges each configuration has and the target configurations of those edges.

\begin{theorem}\label{theorem:correct_encoding}
    Given a CGS $S = \langle k, Q, \Pi, \pi, d, \delta \rangle$, a state $q\in Q$, and an ATL formula $\phi$, let $G$ be the negation-safe EDG rooted in $\langle q, \phi\rangle$ constructed following \cref{fig:dg_p_true_neg,fig:dg_or,fig:dg_next,fig:dg_until,fig:dg_anti_until,fig:dg_partial}. Then $\alpha^G_{\min}(\langle q, \phi\rangle)=1$ if and only if $q\vDash\phi$.
\end{theorem}
The proof of \cref{theorem:correct_encoding} can be found in the extended version of this paper.


%% file: Sections/Model_Checking/Certain_Zero.tex
\paragraph{Certain Zero}\label{sec:certain_zero}

\begin{figure}
    \centering
    \begin{minipage}[c]{35mm}
        \begin{tikzpicture}
            \node (unexplored) at (0, 0) {$\bot$};
            \node (unknown) at (0, 1) {$?$};
            \node (zero) at (-0.7, 2) {0};
            \node (one) at (0.7, 2) {1};
            \draw (unexplored) -- (unknown);
            \draw (unknown) -- (zero);
            \draw (unknown) -- (one);
        \end{tikzpicture}
    \end{minipage}\hspace{0mm}
    \begin{minipage}[c]{0.4\textwidth}
        \caption{Ordering of assignment values for fixed-point computation in the \czero algorithm. The top values are more certain.}
        \label{fig:symb_assign_values_ord}
    \end{minipage}
\end{figure}
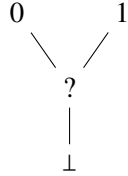

In addition to the global algorithm (repeated application of $F_i$ from \cref{eq:assignment_mono_func}), \thetool also implements a local algorithm heavily inspired by the distributed \czero algorithm by Dalsgaard~et~al.~\cite{dalsgaard}.
A local algorithm specialises in finding the assignment $\alpha_{\min}^G(\langle q, \phi\rangle)$ of a specific configuration $\langle q, \phi\rangle$ and not necessarily the assignment of the whole EDG. This allows the algorithm to terminate early in many cases.
For the \czero algorithm, Dalsgaard~et~al.\ also introduce a symbolic assignment with new assignment values. The value $\bot$ indicates that the configuration is unexplored. The value $?$ (unknown) indicates that the configuration is explored, but its final value has yet to be determined. Lastly, 0 and 1 indicate the final values in the minimum fixed-point assignment $\alpha_{\min}^G$. This induces an ordering as seen in the lattice in \cref{fig:symb_assign_values_ord}. During the algorithm, the assignments of the configurations will rise to more and more certain values, and if the assignment of a configuration ever becomes 0 or 1, we can be certain that its final value is 0 or 1, respectively. This is one of the properties that lead to improved performance, since in contrast to the global algorithm where all configurations are initially assigned 0, we can distinguish an initial 0 and a certain 0. Hence, the \czero algorithm can terminate early if the root configuration is ever assigned either 0 or 1. Otherwise, it terminates when it can no longer raise any assignment values.

%% file: Sections/Model_Checking/Search_Strategies/Search_Strategies.tex
\subsection{Search Strategies}\label{sec:search_strategies}

The \czero algorithm is controlled by a search strategy, which determines in which order the edges are explored and evaluated.
Our search strategies for \thetool include the common breadth-first search (BFS) and depth-first search (DFS) strategies, as well as multiple search strategies based on heuristics. Some of these are discussed in the following subsections. Which strategy is best depends heavily on the shape of the EDG which is determined by the CGS and the ATL formula in question. An evaluation of the strategies can be found in \cref{sec:evaluation}.
The BFS strategy is the default strategy.

\input{Sections/Model_Checking/Search_Strategies/Dependency_Heuristic_Search}
\input{Sections/Model_Checking/Search_Strategies/Linear_Programming_Heuristic_Search}
\input{Sections/Model_Checking/Search_Strategies/Instability_Heuristic_Search}

%% file: Sections/Model_Checking/Search_Strategies/Dependency_Heuristic_Search.tex
\paragraph{Dependency Heuristic Search (DHS)}\label{sec:dependency_heuristic_search}

PageRank~\cite{Page1998PageRank} is an algorithm that was created to estimate the importance of a website based on how many other websites have links to it.
The idea has since been used in other areas, such as graph recommendation systems~\cite{lee20153396} or measuring structural-context similarity with SimRank~\cite{jeh2001simrank}.
Our dependency heuristic search (DHS) uses a similar idea by assuming that configurations with many ingoing edges are important.
Finding the assignments of these, results in more back-propagation, bringing us closer to termination.
In other words, the heuristic focuses on the \textit{trunk} of the EDG where certain assignments are of high value.
Specifically, DHS prioritise edges where the source configuration has a high number of ingoing edges.
That is, if $e$ is an edge with source configuration $c$ then:
\begin{align}
    priority(e) = indegree(source(e)) = |\{ e' \mid (e' = \langle c', T \rangle \in E \land c \in T) \lor e' = \langle c', c \rangle \in N \}|
\end{align}

Edges with the same priority are subject to FIFO ordering.
However, since we do not know the whole graph in advance, we must continuously update the priority of edges when we explore successors of new configurations.
This is only a small overhead with a priority queue data structure.

%% file: Sections/Model_Checking/Search_Strategies/Linear_Programming_Heuristic_Search.tex
\paragraph{Linear Programming Heuristic Search (LPS)}\label{sec:linear_programming_heuristic_search}

In this search strategy, we take advantage of how LCGS states can be represented as vectors, i.e.\ $Q$ is a vector space.
Given an edge with source configuration $\langle q, \phi\rangle$, we transform $\phi$ into a set $\mathcal L_\phi$ of linear constraints, each defined as a pair $\langle \textbf{C},b\rangle$, where $\textbf{C}$ is a matrix and $b$ is a vector. We prioritise edge $e$ if its source configuration $\langle q,\phi\rangle$ has a low estimated distance to satisfaction as given by:
\begin{align}
    dist_{LPS}(\langle q, \phi\rangle)=\min || s - q ||_1 \quad\text{subject to } \textbf{C}s\geq b \quad\text{ where } s\in Q,\;\langle \textbf{C}, b \rangle\in\mathcal{L}_\phi
\end{align}
where $||\cdot||_1$ is the 1-norm, i.e.\ taxicap distance. The following equation is an equivalent linear programming problem:
\begin{align}
    & \min \sum_i x_i \quad\text{subject to } \textbf{C}s\geq b \text{ and } -x_i\leq s_i - q_i \leq x_i \quad \forall i \quad\text{ where } s\in Q,\;\langle \textbf{C}, b \rangle\in\mathcal{L}_\phi
\end{align}

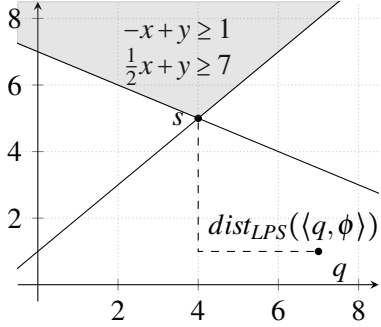
\begin{figure}
    \centering
    \begin{minipage}[c]{0.35\textwidth}
        \begin{tikzpicture}
            \begin{axis}[
              axis lines=middle,
              axis line style={},
              xmin=-0.5,xmax=8.5,ymin=-0.5,ymax=8.5,
              xtick distance=2,
              ytick distance=2,
              grid=major,
              grid style={thin,densely dotted,black!20},
              scale=0.7]
            \addplot [domain=-1:16,samples=2] {x+1};
            \addplot [domain=-1:16,samples=2] {-x/2+7};
            \node[label={330:{$q$}},circle,fill,inner sep=1pt] (q) at (axis cs:7,1) {};
            \node[label={180:{$s$}},circle,fill,inner sep=1pt] (s) at (axis cs:4,5) {};
            \draw[dashed] (q) -| node[above right] {$dist_{LPS}(\langle q, \phi\rangle)$} (s);
            \node[text width=2cm] (cb) at (axis cs:4,7) {$-x+y\geq 1$\\$\frac 1 2 x + y \geq 7$};
            \draw[opacity=0.2,fill=gray] (axis cs: 4,5) -- (axis cs: -0.5,7.25) -- (axis cs: -0.5,8.5) -- (axis cs: 7.5,8.5) -- cycle; 
            \end{axis}
        \end{tikzpicture}
    \end{minipage}\hspace{5mm}
    \begin{minipage}[c]{0.6\textwidth}
        \caption{The distance $dist_{LPS}(\langle q, \phi\rangle)=7$ when $q=\langle 7, 1\rangle\in Q=\R^2$ and $\mathcal{L_\phi}$ corresponds to the constraints $-x+y\geq 1$ and $\frac 1 2 x + y \geq 7$. The state $s$ minimises $||s-q||_1$ while adhering to the constraints.}
        \label{fig:linear_optimise_heuristic}
    \end{minipage}
\end{figure}

A visualisation of the distance $dist_{LPS}$ can be seen on \cref{fig:linear_optimise_heuristic}, where $Q=\R^2$, $q=\langle 7, 1\rangle\in Q$ and
\[
    \mathcal{L}_\phi = \{\langle\begin{bmatrix}-1&1\\\frac{1}{2}&1\end{bmatrix}, \begin{bmatrix}1\\7\end{bmatrix}\rangle\}
\]

Many methods exist for solving linear programming problems.
We use a library called \texttt{minilp}\footnote{\texttt{minilp} crate: \url{https://crates.io/crates/minilp}}.
The LPS strategy assumes value-wise close states are structurally close and works best when the constituent variables of states represent non-categorical data.
The drawback of this method is the computational overhead of linear programming amplified by the potential need to solve multiple linear programming problems for each configuration based on the structure of $\phi$.
We reduce some of this overhead by caching the set $\mathcal L_\phi$ for each $\phi$.

We also support an alternative search strategy involving linear programming. It is called Linear Representative Search (LRS) and it computes the distance described above for the root configuration only. The closest satisfying state $s$ is then saved and edges are prioritised based on the 1-norm distance between $s$ and the state in their source configuration. In other words, we assume that the state $s$ found for the root represents all satisfied states. As a result this search strategy is cheaper than LPS but risks being inaccurate.

%% file: Sections/Model_Checking/Search_Strategies/Instability_Heuristic_Search.tex
\paragraph{Instability Heuristic Search (IHS)}\label{sec:instability_heuristic_search}
The Petri Net model checker \tapaal~\cite{Jensen2016Tapaal} implements the \czero algorithm as well.
Their configurations consist of a marking (a state) and a property, and their default search strategy uses a heuristic that estimates the distance between the marking of the configuration and a marking that satisfies the formula of the configuration.
Our novel instability heuristic is inspired by their heuristic, but we differ by acknowledging that since there are negation edges in the EDG, we may not always be looking for a state that satisfies the formula.
That is, if the state already satisfies the associated formula, we estimate the distance to a state that does not instead.
This guides the search towards configurations where the assignment is \textit{unstable} and thus may have a high influence.
\cref{alg:dist_ihs,alg:bi_dist} find the $dist_{IHS}$ described above for an edge.
The algorithms use an abstract metric $BiDist$ which for state-property pairs finds $\langle\hat t,\hat f\rangle$ where $\hat t$ is the distance to the property being true and $\hat f$ is the distance to the property being false.

\begin{algorithm}[H]
\caption{$dist_{IHS}$}\label{alg:dist_ihs}
\Input{An edge $e\in E$}
\Output{An estimated distance to a different assignment}

\If{$e=\langle c, T\rangle\in E$}{
    $\langle\hat t, \hat f\rangle := \bigsqcap_{\langle q, \phi\rangle\in T} BiDist(q, \phi)$\;
    \Return{$\hat t$ if $\hat t>0$, otherwise $\hat f$}
}
\ElseIf{$e=\langle c, \langle q, \phi\rangle\rangle\in N$}{
    $\langle\hat t,\hat f\rangle := BiDist(q, \phi)$\;
    \Return{$\hat f$ if $\hat f<0$, otherwise $\hat t$}
}
\end{algorithm}

\begin{algorithm}[H]
\caption{$BiDist$}\label{alg:bi_dist}
\Input{A state $q$ and formula $\phi$}
\Output{A bi-distance $\langle\hat t,\hat f\rangle$ describing the instability of $\phi$ in $q$}

\If{$\phi=expr_1\lhd expr_2$ \emph{\bf and} $v=eval(q, expr_1) - eval(q, expr_2)$}{
    \Return{$\langle v, 0 \rangle$ if $v>0$, otherwise $\langle 0, v\rangle$}
}
\ElseIf{$\phi=expr_1\rhd expr_2$ \emph{\bf and} $v = eval(q, expr_2) - eval(q, expr_1)$}{
    \Return{$\langle v, 0 \rangle$ if $v>0$, otherwise $\langle 0, v\rangle$}
}
\ElseIf{$\phi=\neg\phi'$ \emph{\bf and} $\langle\hat t, \hat f\rangle=BiDist( q, \phi')$}{
    \Return{$\langle\hat f, \hat t\rangle$}
}
\ElseIf{$\phi=\phi_1\land\phi_2$}{
    \Return{$BiDist(q,\phi_1)\sqcap BiDist(q,\phi_2)$}
}
\ElseIf{$\phi=\phi_1\lor\phi_2$}{
    \Return{$BiDist(q,\phi_1)\sqcup BiDist(q,\phi_2)$}
}
\ElseIf{$\phi=\enforce{A}\atlnext \phi'$}{
    \Return{$BiDist(q,\phi')$}
}
\ElseIf{$\phi=\enforce{A}\until{\phi_1}{\phi_2}$}{
    \Return{$BiDist(q,\phi_1)\sqcup BiDist(q,\phi_2)$}
}
\ElseIf{$\phi=\despite{A}\until{\phi_1}{\phi_2}$}{
    \Return{$BiDist(q,\phi_1)\sqcup BiDist(q,\phi_2)$}
}
\tcp{where $\lhd\in\{<,\leq\}, \langle\hat t_1,\hat f_1\rangle\sqcap\langle\hat t_2,\hat f_2\rangle=\langle\hat t_1 +\hat t_2,\min\{\hat f_1, \hat f_2\}\rangle$ \\
$\quad\quad\quad \rhd\in\{>,\geq\}, \langle\hat t_1,\hat f_1\rangle\sqcup\langle\hat t_2,\hat f_2\rangle=\langle\min\{\hat t_1, \hat t_2\},\hat f_1 +\hat f_2\rangle$
}
\end{algorithm}

%% file: Sections/Tool_Overview/Tool_Overview.tex
\section{Tool Overview}\label{sec:tool_overview}
\thetool(\url{https://github.com/d702e20/CGAAL}) is written in Rust and consists of a command-line interface and a verification engine.
The primary feature of \thetool is the verification of ATL properties for CGSs.
Verification is either done with a global algorithm or local algorithm, both of which are described in \cref{sec:model_checking}.
If requested, a partial strategy witness can also be computed, which instructs how the given players must play to satisfy the given property.
Another feature converts the model into \texttt{dot} graph format such that it can be visually rendered with Graphviz.\footnote{Graphviz: \url{https://graphviz.org/}}

To model concurrent game structures for \thetool, we designed a declarative Language for Concurrent Game Structure, called LCGS. In LCGS a CGS is defined with a series of declarations such that it is possible to find the successors of any state from the declarations alone. The language acts as an abstract representation of the CGS and allows us to save memory at the small cost of having to evaluate the expressions of declarations whenever successors are explored. The syntax of the language is inspired by \prism-lang~\cite{Parker2011Prism} used by the \prism model checker to model stochastic multi-player games with rewards. However, LCGS differs in multiple ways. For instance, LCGS has player templates instead of modules, and templates have no effect unless there exists an instance of it.
Additionally, synchronisations affecting the internal state of a player are much easier to declare.

\paragraph{Example Use}\label{sec:cgaal_example_use}
As an example, we want to check if a cowboy can guarantee to stay alive in a three-way Mexican standoff. The standoff is simulated in rounds and in each round a cowboy can choose to wait, shoot the cowboy to the right, or shoot the cowboy to the left. If a cowboy is hit by two bullets, he dies.

\begin{lstlisting}[caption={LCGS implementation of a Mexican standoff},label={code:mexican_standoff_lcgs}]
const max_health = 2;

template cowboy

    // How many bullets a cowboy can be hit by before dying
    health : [0 .. max_health] init max_health;
    health' = max(health - opp_right.shoot_left - opp_left.shoot_right, 0);
    
    // A proposition used by ATL formulae
    label alive = health > 0;

    // The actions available to each cowboy
    [wait] 1;
    [shoot_right] health > 0 && opp_right.health > 0;
    [shoot_left] health > 0 && opp_left.health > 0;

endtemplate

// The three cowboys in the Mexican standoff
player billy = cowboy [opp_right=clayton, opp_left=jesse];
player clayton = cowboy [opp_right=jesse, opp_left=billy];
player jesse = cowboy [opp_right=billy, opp_left=clayton];
\end{lstlisting}

We model this scenario in \cref{code:mexican_standoff_lcgs}. The cowboy template is declared on lines 3-17. Herein, on line 6, we define that each cowboy can be hit by two bullets before being incapacitated, but we make this number a constant on line 1 so it is easy to change. Each cowboy has their own health variable and the combination of the values of these variables makes up a state. Line 7 defines how the health of a cowboy is updated in each transition. The value of \texttt{opp\_right.shoot\_left} and \texttt{opp\_left.shoot\_right} are 1 if this cowboy was shot by the given opponent to the right or left, respectively, otherwise 0. On line 10 we define a label called alive, which is a proposition that is true if the cowboy has more than zero health. Lines 13-15 define the actions of the cowboy template, and the expression to the right of the name is a condition defining in which states the action is available. As can be seen, a cowboy can always wait, but only shoot if they and their target are alive. Lastly, we declare three instances of the cowboy template on lines 20-22. We define the right and left opponent in the relabelling function after the name of the template. Any identifier in a template can be relabelled to another expression as long as the result is syntactically and semantically correct.

The ATL formula $\enforce{\texttt{billy}}\invariantly \texttt{billy.alive}$ is satisfied if the cowboy \texttt{billy} has a strategy to stay alive. We can now run \thetool with the following command: \texttt{./cgaal solver -m standoff.lcgs -f billy-can-stay-alive.atl}, and \thetool will tell us that the property is not satisfied. Billy has no strategy that can guarantee his survival.

%% file: Sections/Evaluation/Evaluation.tex
    \section{Evaluation}\label{sec:evaluation}
To evaluate our tool, we run several experiments.
In our experiments we compare our global algorithm, our local algorithm using our various search strategies, and the established tool \prism-games.
We use several different concurrent game case studies. Some of these are \prism-games case studies adapted to LCGS for \thetool and determinised if needed.
Others are well-known algorithms and games constructed during the development of \thetool.
We make the \prism-lang and LCGS implementations of models as identical as possible, such that the state spaces are comparable.
Some of the concurrent games and related ATL formulae used in the experiments are presented below.
Queries marked with $\top$ (resp. $\bot$) are satisfied (resp. not satisfied), and queries marked with $\dagger$ may terminate early as we are not required to compute the entire fixed-point:
\begin{itemize}
    \item \textbf{Mexican-Standoff}:
        In this model $N$ cowboys stand in a circle, each with a revolver.
        At each moment they can choose to shoot another cowboy or do nothing.
        If a cowboy is hit by $B$ bullets, they are incapacitated.
        We run the following queries:
        \vspace{-2mm}
        \begin{align*}
            &\testMSone N B=\enforce{p_1}\invariantly p_1\_alive\\
            &\testMStwo N B=\enforce{p_1}\eventually \neg p_1\_alive\\
            &\testMSthree N B=\enforce{\{p_i\in\Sigma\mid 1 \equiv i \mod 2\}}\invariantly p_1\_alive
        \end{align*}
    \item \textbf{Gossiping Girls}:
        In this model, $N$ nodes each know a piece of information.
        The nodes can exchange information by calling each other.
        Our experiments include both a fully connected network of nodes and a circular network of nodes.
        We run the following queries:
        \vspace{-2mm}
        \begin{align*}
            &\testGPone N=\enforce{\Sigma}\until{calls\leq 10}{all\_girls\_know\_all}\\
            &\testGPtwo N=\enforce{\Sigma}\until{calls\leq 10}{only\_p_1\_knows\_all}\\
            &\testGPthree N=\enforce{\Sigma}\until{calls\leq 10}{p_1\_knows\_all}\\
            &\testGPfour N=\enforce{\Sigma}\eventually calls>10\land all\_girls\_know\_all\\
            &\testGPfive N=\enforce{\emptyset}\eventually calls> 10\\
            &\testGPsix N=\enforce{p_1}\until{calls\leq 10}{p_1\_knows\_all}\\
            &\testGPseven N=\enforce{\Sigma\setminus\{p_1\}}\eventually everyone\_except\_p_1\_knows\_all
        \end{align*}
    \item \textbf{Robot Coordination}:
        In this model, four robots move orthogonally on a $N$ by $N$ grid.
        Each robot needs to reach the opposite corner of where they start, but crashing into each other is fatal.
        We run the following queries:
        \vspace{-2mm}
        \begin{align*}
            &\testRCone N=\enforce{p_1,p_2,p_3}\eventually p_1\_at\_target\land p_2\_at\_target\\
            &\testRCtwo N=all\_robots\_at\_home\\
            &\testRCthree N=\enforce{p_1,p_2,p_3,p_4}\eventually all\_robots\_at\_targets\\
            &\testRCfour N=\enforce{p_1,p_3,p_4}\eventually p_1\_at\_target\land\neg p_1\_crashed
        \end{align*}
\end{itemize}
Each experiment is run with a time limit of two hours, allocated 128 GB of memory, and has 32 cores available regardless of how many worker threads \thetool may spawn.
All experiments are run on several identical AMD EPYC 7642 based servers, allowing only one experiment per node at a time to reduce noise in the results.

\input{Sections/Evaluation/Results}

%% file: Sections/Evaluation/Results.tex
\paragraph{Results}\label{sec:results}

\begin{figure}
    \centering
    \begin{subfigure}[t]{0.49\textwidth}
        \includegraphics[width=\textwidth]{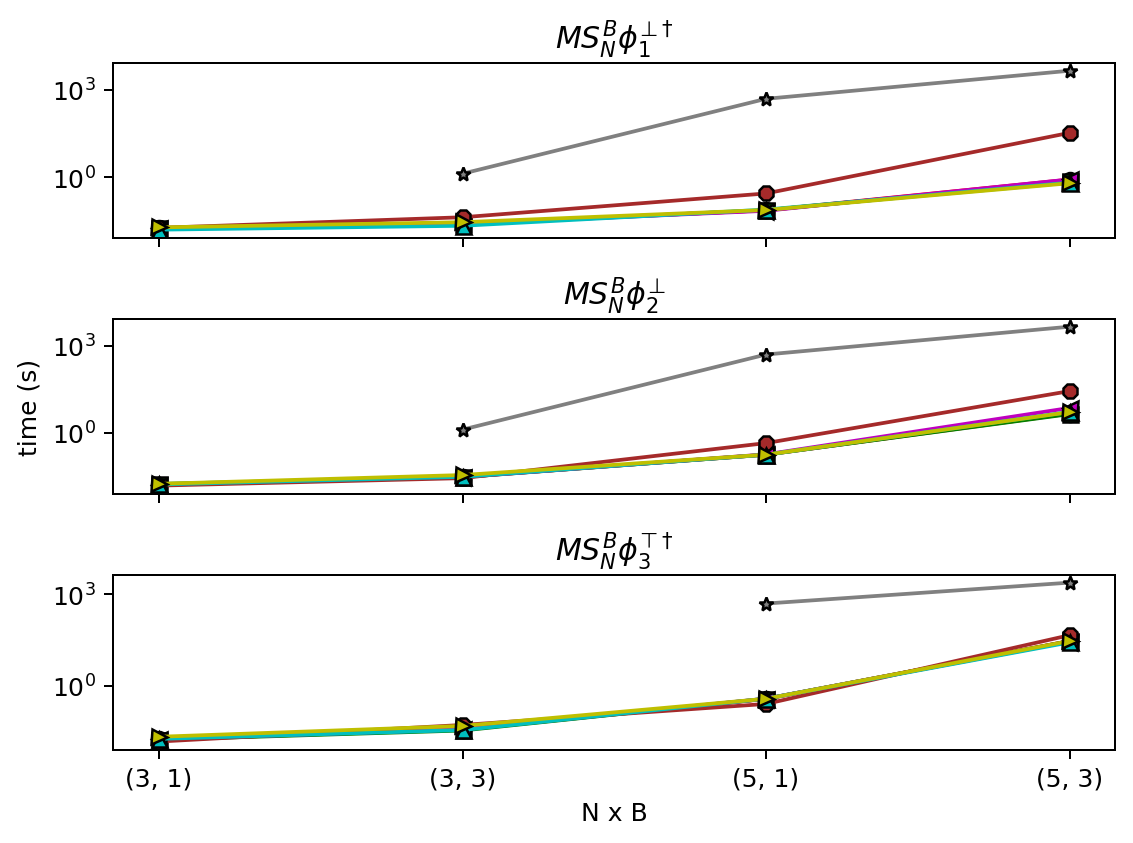}
        \caption{Mexican-standoff: horizontal axis is $(N, B)$ tuples with $N$ being the number of cowboys in the model and $B$ the number of bullets the cowboys can be hit by.}
        \label{fig:mex-result}
    \end{subfigure}
    \hfill
    \begin{subfigure}[t]{0.49\textwidth}
        \includegraphics[width=\textwidth]{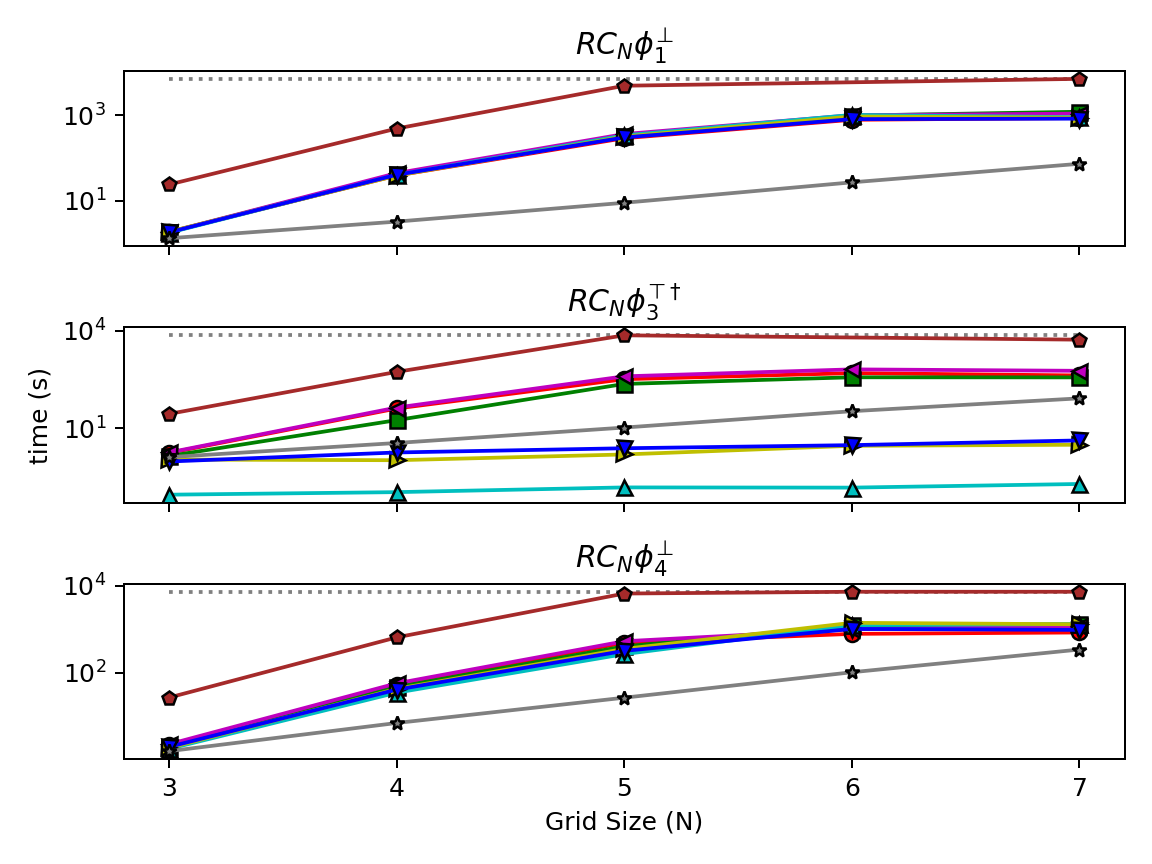}
        \caption{Robot Coordination: horizontal axis is the size of the grid the robots are to manoeuvre on.}
        \label{fig:rc-result}
    \end{subfigure}
    \\
    \begin{subfigure}[t]{0.49\textwidth}
        \includegraphics[width=\textwidth]{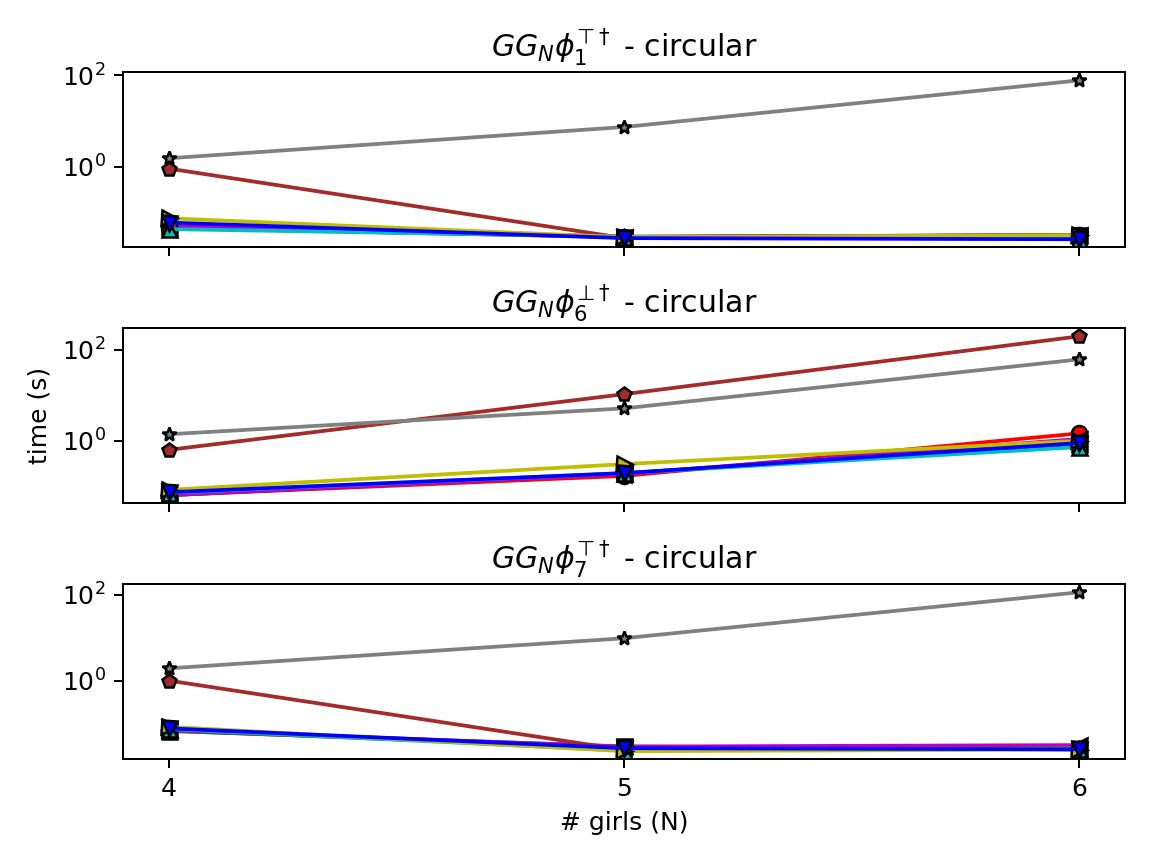}
        \caption{Gossiping girls in a circular topology: horizontal axis being the number of girls in the circle.}
        \label{fig:gg-result}
    \end{subfigure}
    \hfill
    \begin{subfigure}[t]{0.49\textwidth}
        \includegraphics[width=\textwidth]{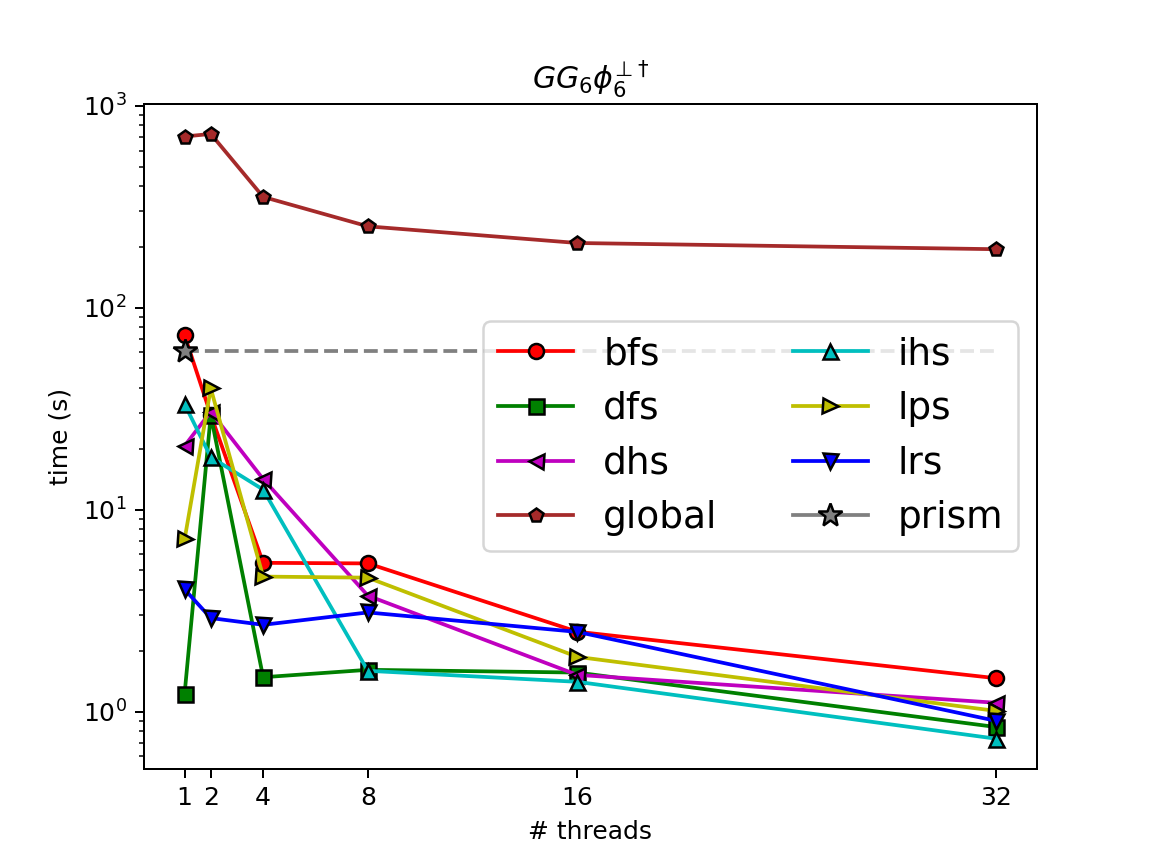}
        \caption{Gossiping girls with six girls in a circular topology: horizontal axis being the number of compute threads used in our distributed algorithm. Note that \prism was always run with just one thread.}
        \label{fig:gg-threads}
    \end{subfigure}
    \caption{Select experimental results. \thetool-results in \cref{fig:mex-result,fig:rc-result,fig:gg-result} show the best result for each search strategy when varying between 1 and 32 threads.}
\end{figure}

Here we provide a select set of experimental results that roughly exemplify our findings in general.
For a more thorough set of results along with all of our experimental data and the Python script processing it, see the git repository \href{https://github.com/d702e20/evaluation-results/}{https://github.com/d702e20/evaluation-results/}.

We find that the local on-the-fly algorithm is often one order of magnitude faster than the global algorithm regardless of the search strategy. However, the global algorithm can compete with and sometimes outperforms the local algorithm in cases where the local algorithm cannot terminate early.
This matches observations made by A.~Dalsgaard et~al.~\cite{dalsgaard} and M.~C.~Jensen et~al.~\cite{probabilisticCTL}.

\prism-games is a stochastic model checker and checks queries involving both probability and rewards, capabilities which are irrelevant in our test cases. Therefore, in many cases, \thetool's local algorithm is two orders of magnitude faster than \prism-games, typically when the model has many synchronisations that affect internal states of modules such as in Mexican Standoff and Gossiping Girls as seen in \cref{fig:mex-result,fig:gg-result}. Such synchronisations also require a high amount of \prism-lang code, while being easily expressed in LCGS. However, there are a few cases where \prism-games is faster than \thetool, e.g.\ \testRCone{N} and \testRCfour{N} as seen in \cref{fig:rc-result}. Here the local algorithm cannot terminate early.

The choice of search strategy sometimes has a discernible impact on the performance of the local algorithm.
Which search strategy is best varies from case to case and no definitive best strategy.
The BFS strategy does not work well when the local algorithm can terminate early based on information found multiple moves into the model, but its low overhead is beneficial when early termination is not possible.
The IHS strategy is often a good choice when early termination is possible.
The LPS strategy is often significantly slower than the others, which is unsurprising given its overhead of solving linear programming problems.
In models where differences in the states correspond to a distance in space, such as in robot coordination, the LPS strategy performs notably better and is even faster than both BFS and DFS on \testRCthree{4}.
The lightweight version, LRS, which only solves the linear problem once, matches LPS in these cases but is also generally better due to its lower overhead.

A general trend in our experiments is that we see an increase in the execution speed of our distributed implementation as we increase the number of compute threads available. 
The only times we do not see a speed-up with an increased number of compute threads are when the models are small enough such that the overhead of managing the additional threads is significant.
Similarly, a single thread performs better than a few threads in many cases, since there is no communication overhead with only one compute thread.
In general, depending on the search strategy employed, we see an improvement on a scale of one to two orders of magnitude when increasing the number of compute threads as exemplified in \cref{fig:gg-threads}.

%% file: Sections/Conclusion.tex
\section{Conclusion}\label{sec:conclusion}

In this paper, we present \thetool, our model checker of alternating-time temporal logic properties in concurrent games.
\thetool checks such properties by encoding the problem as an extended dependency graph and then computes the satisfaction relation using the distributed local on-the-fly \czero algorithm by Dalsgaard~et~al.~\cite{dalsgaard}.
We provide multiple novel search strategies for the algorithm and allow concurrent games to be expressed in our language LCGS.
Our experiments show that the local on-the-fly algorithm outperforms the global algorithm in the majority of cases.
We also find that \thetool outperforms the state-of-the-art tool \prism-games by being up to two orders of magnitude faster, especially in models where synchronisations affect the internal state of modules and whenever we are not required to compute the entire fixed point. However, this comparison is unfair, since our test only uses a fraction of \prism-games feature set.

\thetool is still in early development and much work is needed before it competes with \prism in terms of capabilities. However, dependency graphs have been used for encoding various model-checking problems, and we intend to incorporate these techniques into \thetool.